\documentclass{svproc}
\usepackage{url}

\usepackage{xfrac,amssymb,amsfonts,amsmath}
\usepackage{tikz,pgfplots}
\usetikzlibrary{positioning}
\usetikzlibrary{matrix}

\begin{document}
\mainmatter              % start of a contribution
\title{Point transformations: exact solutions of the quantum time-dependent mass nonstationary oscillator}
\titlerunning{Point trans.: exact sols. of the quantum T.D. mass nonstationary oscillator}  % abbreviated title (for running head)
\author{Kevin Zelaya\inst{1} \and V\'eronique Hussin\inst{1,2}}
\authorrunning{K. Zelaya and V. Hussin} % abbreviated author list (for running head)
\tocauthor{Kevin Zelaya and V\'eronique Hussin}
\institute{Centre de Recherches Mathematiques, Universit\'e de Montr\'eal,\\
Montr\'eal H3C 3J7, QC, Canada,\\
\email{zelayame@crm.umontreal.ca},
\and D\'epartment de Math\'ematiques et de Statistique, Universit\'e de Montr\'eal,\\ Montr\'eal H3C 3J7, QC, Canada\\
\email{hussin@dms.umontreal.ca}}

\maketitle              % typeset the title of the contribution

\begin{abstract}
In this note we address the exact solutions of a time-dependent Hamiltonian composed by an oscillator-like interaction with both a frequency and a mass term that depend on time. The latter is achieved by constructing the appropriate point transformation such that it deforms the Schr\"odinger equation of a stationary oscillator into the one of the time-dependent model. Thus, the solutions of the latter can be seen as deformations of the well known solutions of the stationary oscillator, and thus an orthogonal set of solutions can be determined in a straightforward way. The latter is possible since the inner product structure is preserved by the point transformation. Also, any invariant operator of the stationary oscillator is transformed into an invariant of the time-dependent model. This property leads to a straightforward way to determine constants of motion without requiring to use ansatz.
% We would like to encourage you to list your keywords within
% the abstract section using the \keywords{...} command.
\end{abstract}
\section{Introduction}
The dynamics of non-relativistic quantum systems is determined through the solutions of the Schr\"odinger equation. In the latter, the information of the system under consideration is coded in the Hamiltonian operator. Most of the physical systems of interest are stationary and are described by time-independent Hamiltonians. On the other hand, for open systems, time-dependent Hamiltonians are required to provide an accurate description. Physical applications are found in electromagnetic traps of particles~\cite{Pri83,Cum86,Mih09}, in which external time-dependent electric and magnetic fields allow the confinement of particles~\cite{Gla92}. In such case we describe the respective Hamiltonian through a \textit{parametric oscillator} potential, also known as \textit{nonstationary oscillator}, which consists of an oscillator-like interaction with a frequency that varies in time. Exact solutions were studied in detail by Lewis and Riesenfeld in the classical and quantum cases~\cite{Lew68,Lew69}. Given that the Hamiltonian has an explicit dependence on time, an eigenvalue equation associated with Hamiltonian is not longer feasible and the existence of an orthonormal set of solutions can not be taken by granted, as is customary for the stationary quantum oscillator. Nevertheless, Lewis and Riesenfeld introduced an approach in which a nonstationary eigenvalue equation can be still found once the appropriate constant of motion is determined~\cite{Lew69}. With the latter, solutions to the Schr\"odinger equation are found by adding the appropriate time-dependent complex-phase to the nonstationary eigenfunctions. The constant of motion, or \textit{invariant operator} $\hat{I}(t)$, is usually imposed as an ansatz and determined from the condition $[i\partial/\partial t-\hat{H}(t),\hat{I}(t)]=0$. Such approach has been applied successfully to other time-dependent models as well~\cite{Dod98}. 

In this note we address the solutions of the nonstationary oscillator with time-dependent mass. To this end we consider the method of point transformations~\cite{Zel19c}. The latter is achieved by deforming the well known Schr\"odinger equation of the stationary oscillator into the one of the time-dependent model. This method leads to a straightforward way to obtain the solutions of the time-dependent model as deformations of the stationary oscillator. Remarkably, the point transformation preserves the first integrals, this means that the constants of motion and the spectral properties for the time-dependent model are inherited from the stationary oscillator, without requiring to impose any anstaz~\cite{Lew69}.

\section{Nonstationary oscillator with time-dependent mass}
Let us first consider the quantum harmonic oscillator, defined through the Hamiltonian
\begin{equation}
\hat{H}_{osc}=\frac{\hat{p}_{y}^{2}}{2}+\frac{\hat{y}^{2}}{2}, 
\label{eq:INT0}
\end{equation}
where $\hat{y}$ and $\hat{p_{y}}$ stand for the canonical position and momentum operators, $[\hat{y},\hat{p}_{y}] = i$. With the latter, the Schr\"odinger equation, represented in the spatial coordinate `$y$', reads as
\begin{equation}
i\frac{\partial\Psi}{\partial \tau} = - \frac{1}{2}\frac{\partial^{2}\Psi}{\partial y^2} + \frac{y^2}{2} \Psi,
\label{eq:INT1}
\end{equation}
where $\tau$ is the time-parameter, the momentum operator was represented as $\hat{p}_{y}=-i\frac{\partial}{\partial y}$ and $\Psi(y,\tau)=\langle y \vert \Psi(\tau)\rangle$ is the respective wave function. Given that $\hat{H}_{osc}$ is time-independent, the solutions of~(\ref{eq:INT1}) can be easily computed by using the separation of variables $\Psi(y,\tau)=e^{-i E \tau}\Phi(y)$, where $\Phi(y)= \langle y \vert\Phi \rangle$ fulfills the eigenvalue equation
\begin{equation}
-\frac{1}{2}\frac{d^{2}\Phi}{dy^2} + \frac{y^2}{2} \Phi = E \Phi.
\label{eq:INT2-2}
\end{equation}
A set of physical solutions $\{ \Phi_{n}(y)\}_{n=0}^{\infty}$ is determined with aid of the finite-norm condition $\vert\vert \vert\Phi_{n}\rangle\vert\vert^{2}=\langle \Phi_{n}\vert\Phi_{n}\rangle<\infty$, where the inner product of two eigenfunctions $\Phi_{(1)}(y)$ and $\Phi_{(2)}(y)$ is defined through
\begin{equation}
\langle \Phi_{(2)}\vert\Phi_{(1)}\rangle=\int_{-\infty}^{\infty}dy\,\Phi_{(2)}^{*}(y)\Phi_{(1)}(y) \, .
\label{eq:INT2-3}
\end{equation}
The spectral information of the harmonic oscillator is then given by
\begin{equation}
\Phi_{n}(y)=\sqrt{\frac{1}{2^{n}n!\sqrt{\pi}}} \, e^{-\frac{y^{2}}{2}}H_{n}\left( y \right), \quad E_{n}=(n+1/2),
\label{eq:INT3}
\end{equation}
where $H_n(z)$ are the Hermite Polynomials \cite{Olv10}. The set of eigenfunctions is orthonormal, $\langle \Phi_{m}\vert\Phi_{n}\rangle=\delta_{n,m}$, and it generates the space ${\cal H} = \mbox{span} \{\vert\Phi_{n}\rangle\}_{n=0}^{\infty}$. 

Now, we introduce the nonstationary oscillator with time-dependent mass, defined in terms of the canonical position and momentum operators $\hat{x}$ and $\hat{p}_{x}$, respectively, together with the time parameter $t$ through the time-dependent Hamiltonian
\begin{equation}
\hat{H}(t)=\frac{1}{2m(t)}\hat{p}^{2}_{x}+\frac{1}{2}m(t)\Omega^{2}(t)\hat{x}^{2}+F(t)\hat{x}+V_{0}(t),
\label{eq:POM1}
\end{equation}
where $m(t)$ is the time-dependent mass, $\Omega^{2}(t)$ the time-dependent frequency, $F(t)$ an external driving force and $V_{0}(t)$ a zero-point energy term. The wave functions $\psi(x,t)=\langle x \vert \psi(t)\rangle$ associated with the Hamiltonian~(\ref{eq:POM1}) are thus computed from the Schr\"odinger equation
\begin{equation}
i\frac{\partial\psi}{\partial t} = -\frac{1}{2m(t)}\frac{\partial^2\psi}{\partial x^2} \psi + \frac{1}{2}m(t)\Omega^{2}(t)x^2 \psi +F(t)x \psi +V_{0}(t) \psi ,
\label{eq:INT4}
\end{equation}
In the sequel, we address the solutions of~(\ref{eq:INT4}) by constructing the point transformation.
\section{Point transformation}
In order to transform the Schr\"odinger equation of the stationary oscillator~(\ref{eq:INT1}) into the one of the nonstationary oscillator with time-dependent mass~(\ref{eq:INT4}), let us analyze the relationships between the elements of the set $\{ y, \tau, \Psi \}$ and those of the set $\{ x, t, \psi \}$ of the form~\cite{Ste93}
\begin{equation}
y=y(x,t), \quad  \tau =  \tau (x,t), \quad \Psi= \Psi (y(x,t), \tau (x,t))=G(x,t;\psi(x,t)).
\label{eq:INT5}
\end{equation}
The dependence of $\Psi$ on $x$ and $t$ is implicit, so, we have introduced the function $G$ as a reparametrization that allows to rewrite $\Psi$ as an explicit function of $x$, $t$ and $\psi$. In this way, we have at hand a mechanism to map any solution of~(\ref{eq:INT1}) into a solution of~(\ref{eq:INT4}). The explicit form of the relationships in~(\ref{eq:INT5}) is determined through the total derivatives $\frac{d\Psi}{dx}$, $\tfrac{d\Psi}{dt}$ and $\frac{d^{2}\Psi}{dx^{2}}$. It allows to find relationships between the partial derivative of the initial and final models, leading to the forms
\begin{equation}
\frac{\partial\Psi}{\partial \tau}=G_{1}\left( x,t;\psi,\frac{\partial\psi}{\partial x},\frac{\partial\psi}{\partial t} \right) \, , \quad \frac{\partial^{2}\Psi}{\partial y^{2}}=G_{2}\left( x,t;\psi,\frac{\partial\psi}{\partial x},\frac{\partial^{2}\psi}{\partial x^{2}},\frac{\partial\psi}{\partial t} \right) \, .
\label{eq:INT8}
\end{equation}
The latter leads in general to nonlinear terms, but the conditions~\cite{Zel19c}
\begin{equation}
\Psi=G(x,t;\psi)=A(x,t)\psi \, , \quad \tau=\tau(t) \, ,
\label{eq:INT9}
\end{equation}
allow to remove such nonlinearities. With~(\ref{eq:INT9}), and after some calculations, it can be shown that the relationships in~(\ref{eq:INT8}) are written as
\begin{equation}
\begin{aligned}
& \Psi_{\tau} = \frac{A}{\tau_{t}}\left[ - \frac{y_{t}}{y_{x}} \psi_x +  \psi_t+\left( \frac{A_{t}}{A}-\frac{y_{t}}{y_{x}} \frac{A_{x}}{A}\right) \psi \right], \\[0.5ex]
& \Psi_{y,y} = \frac{A}{y_{x}^{2}}\left[ \psi_{x,x}+\left( 2 \frac{A_{x}}{A} - \frac{y_{xx}}{y_{x}} \right) \psi_x + \left(\frac{A_{xx}}{A}-\frac{y_{xx} }{y_{x}}\frac{A_{x}}{A} \right)\psi \right],
\end{aligned}
\label{eq:INT10}
\end{equation}
where the subindex notation denotes partial derivatives, $f_u = \frac{\partial f}{\partial u}$. The substitution of~(\ref{eq:INT9}) and~(\ref{eq:INT10}) into~(\ref{eq:INT1}) leads, after some arrangements, to
\begin{equation}
i \psi_t=-\frac{1}{2}\frac{\tau_{t}}{y_{x}^{2}} \psi_{x,x} + B(x,t) \psi_x + V(x,t)\psi,
\label{eq:INT11}
\end{equation}
with
\begin{equation}
\begin{aligned}
& B(x,t)=i\frac{y_{t}}{y_{x}}-\frac{1}{2}\frac{\tau_{t}}{y_{x}^{2}}\left( 2\frac{A_{x}}{A}-\frac{y_{xx}}{y_{x}} \right) ,\\[1ex]
& V(x,t)=-i\left(\frac{A_{t}}{A}-\frac
{y_{t}}{y_{x}}\frac{A_{x}}{A} \right)-\frac{1}{2}\frac{\tau_{t}}{y_{x}^{2}}\left( \frac{A_{xx}}{A}-\frac{y_{xx}}{y_{x}}\frac{A_{x}}{A} \right)+\frac{\tau_{t}}{2} y^{2}(x,t).
\end{aligned}
\label{eq:INT12}
\end{equation}
Given that~(\ref{eq:INT11}) must be of the form~(\ref{eq:INT4}), we impose the conditions
\begin{equation}
\frac{\tau_{t}}{y_{x}^{2}}=\frac{1}{m(t)}, \quad B(x,t)=0.
\label{eq:INT13}
\end{equation}
To simplify the calculations, it is convenient to introduce the real-valued functions $\mu(t)$ and $\sigma(t)$ such that $\tau_{t}=\sigma^{-2}(t)$ and $m(t)=\mu^{2}(t)$. From the first condition in~(\ref{eq:INT13}) we get
\begin{equation}
\tau (t)=\int^{t}\frac{dt'}{\sigma^{2}(t')}, \quad y(x,t)=\frac{\mu(t)x+\gamma(t)}{\sigma(t)},
\label{eq:INT14}
\end{equation}
where the real-valued function $\gamma(t)$ results from the integration with respect to $x$. From $B(x,t)=0$ we obtain $A(x,t)$ as
\begin{equation}
A(x,t)=\exp\left[ i\frac{\mu}{\sigma}\left(\frac{\mathcal{W}_{\mu}}{2}x^{2}+\mathcal{W}_{\gamma}x+\eta\right)\right],
\label{eq:INT15}
\end{equation}
where $\eta(t)$ is a complex-valued function resulting from the integration with respect to $x$ and
\begin{equation}
\mathcal{W}_{\mu}(t)=\sigma\dot{\mu}-\dot{\sigma}\mu, \quad \mathcal{W}_{\gamma}(t)=\sigma\dot{\gamma}-\dot{\sigma}\gamma, 
\end{equation}
with $\dot f = \frac{df}{dt}$. With~(\ref{eq:INT15}), the new time-dependent potential $V(x,t)$ in~(\ref{eq:INT12}) takes the form
\begin{equation}
\begin{aligned}
&V(x,t) = \frac{\mu^{2}}{2}\left(\frac{\dot{\mathcal{W}}_{\mu}}{\mu\sigma}+\frac{1}{\sigma^{4}}\right)x^{2}+\mu\left(\frac{\dot{\mathcal{W}}_{\gamma}}{\sigma}+\frac{\gamma}{\sigma^{4}}\right)x+V_{0}(t) \, , \\
&V_{0}(t)=\frac{\mathcal{W}_{\mu}\xi}{\sigma^{2}}+\frac{\mu\dot{\xi}}{\sigma}-\frac{\mathcal{W}_{\gamma}^{2}}{2\sigma^{2}} + \frac{\gamma^{2}}{2\sigma^{4}} -i\frac{\mathcal{W}_{\mu}}{2\mu\sigma}.
\label{eq:INT16}
\end{aligned}
\end{equation}
After comparing~(\ref{eq:INT16}) with the potential energy term in~(\ref{eq:INT4}) we obtain a system of equations for $\sigma$, $\gamma$ and $\eta$ which, without loss of generality, can be reduced to quadratures by considering\footnote{For $V_{0}(t)\neq 0$, the solutions are just modified by adding a global complex-phase, for details see App.~B of~\cite{Zel19c}.} $V_0(t)=0$. We thus have
\begin{equation}
\ddot{\sigma}+\left(\Omega^{2}-\frac{\ddot{\mu}}{\mu} \right)\sigma=\frac{1}{\sigma^{3}}, \quad \ddot{\gamma}+\left( \Omega^{2}-\frac{\ddot{\mu}}{\mu} \right)\gamma=\frac{F}{\mu}, \quad \frac{\mu}{\sigma}\eta=\xi-\frac{i}{2}\ln\frac{\sigma}{\mu},
\label{eq:INT17}
\end{equation}
where the real-valued function $\xi(t)$ is given by
\begin{equation}
\xi(t)=\frac{\gamma \mathcal{W}_{\gamma}}{2\sigma}-\frac{1}{2}\int^{t}dt'\frac{F(t')\gamma(t')}{\mu(t')}.
\end{equation}
From~(\ref{eq:INT17}) it follows that $\sigma(t)$ satisfies the Ermakov equation~\cite{Erm08}, whose solutions are well known in the literature~\cite{Erm08,Ros15,Bla18}. In general, for a set of nonnegative parameters $\{a,b,c\}$ we have~\cite{Erm08}
\begin{equation}
\sigma(t)= \left[ a q_1^2(t)+ b q_1(t)q_2(t)+c q_2^2(t) \right]^{1/2}, \quad b^2-4ac=-\frac{4}{W_{0}^2},
\label{eq:OSC7}
\end{equation}
where $q_{1}$ and $q_{2}$ are two linearly independent real solutions of the linear equation 
\begin{equation}
\ddot{q}_{1,2}+\left( \Omega^{2}-\frac{\ddot{\mu}}{\mu} \right)q_{1,2}=0 \, ,
\label{eq:LIN}
\end{equation}
and the Wronskian $W(q_1,q_2) =W_0$ is a constant. The constrain in the constants $a$, $b$, $c$ ensures that $\sigma >0$ at any time~\cite{Ros15,Bla18}. Thus, the transformed coordinate $y(x,t)$ and time parameter $\tau(t)$ are free of singularities at any time. Notice that~\eqref{eq:OSC7} corresponds to the classical equation of motion of the parametric oscillator~\cite{Lew68,Gla92}. On the other hand, $\gamma(t)$ is solution to the classical parametric oscillator subjected to a driving force $F(t)$. Ine general, $\gamma$ can be expressed as the sum of the homogeneous solution $\gamma_h = \gamma_{1} q_{1} + \gamma_{2} q_{2}$ and the particular solution $\gamma_{p}(t)$, where the real constants $\gamma_{1,2}$ are fixed according to the initial conditions and the function $\gamma_{p}(t)$ is determined once the driving force $F(t)$ has been specified. Moreover, the function $\tau$ introduced in~(\ref{eq:INT14}) can be rewritten in terms of $q_1$ and $q_2$ as well, leading to~\cite{Bla18}
\begin{equation}
\tau (t)=\int^{t}\frac{dt'}{\sigma^{2}(t')}=\arctan\left[ \frac{W_0}{2}\left( b+2c\frac{q_2}{q_1} \right) \right].
\label{eq:OSC8}
\end{equation}

From~(\ref{eq:INT9}) and with the functions $\sigma$, $\gamma$ and $\tau$ already identified, the solutions to the Schr\"odinger equation~(\ref{eq:INT4}) are simply given in terms of the solutions of the stationary oscillator $\Psi(y,\tau)$ as
\begin{equation}
\psi(x,t)=\exp\left[ -i\frac{\mu}{\sigma}\left(\frac{\mathcal{W}_{\mu}}{2}x^{2}+\mathcal{W}_{\gamma}x+\xi\right)\right] \sqrt{\frac{\mu}{\sigma}}\Psi(y(x,t), \tau(t)) \, .
\label{eq:INN1}
\end{equation}
That is, the solutions of the nonstationary oscillator with time-dependent mass $\hat{H}(t)$ can be seen as a mere deformation of the solutions of the stationary oscillator, provided by the appropriate point transformation. From the latter, it is natural to ask whether the structure of the inner product of the new solutions $\psi(x,t)$ is deformed as well. To this end, let us consider a pair of arbitrary solutions of the stationary oscillator, $\Psi_{(1)}(y, \tau)$ and $\Psi_{(2)}(y, \tau)$. Straightforward calculations shows that
\begin{multline}
\langle\Psi_{(2)}(\tau)\vert \Psi_{(1)}(\tau)\rangle=\int^{\infty}_{-\infty}dy \, \Psi_{(2)}{}^{*}(y, \tau)\Psi_{(1)}(y, \tau) \\
= \int^{\infty}_{-\infty}dx \, \psi_{(2)}{}^{*}(x,t)\psi_{(1)}(x,t)=\langle \psi_{(2)}(t)\vert\psi_{(1)}(t)\rangle,
\label{eq:INN2}
\end{multline}
that is, the point transformation preserves the structure of the inner product. 
\subsection{Orthogonal set of solutions and the related spectral problem}
\label{ortogonal}
With the transformation rule~(\ref{eq:INN1}), we can find an orthogonal  set of solutions for~(\ref{eq:INT4}). From the preservation of the inner product~(\ref{eq:INN2}), it is natural to consider the orthogonal solutions of the stationary oscillator $\Psi_{n}(y,\tau)$, where $\Psi_{n}(y,\tau)=e^{-i(n+1/2)\tau}\Phi_{n}(y)$ and $\Phi_{n}(y)$ is given in~(\ref{eq:INT3}). Hence, the orthogonal set of solutions $\{ \Psi_{n} (y,\tau) \}_{n=0}^{\infty}$ is mapped into an orthogonal set $\{ \psi_{n}(x,t) \}_{n=0}^{\infty}$, where
\begin{equation}
\psi_{n}(x,t)=e^{-i (n+1/2) \tau(t)} \varphi_{n}(x,t) ,
\label{eq:INN2-1}
\end{equation}
with
\begin{equation}
\begin{alignedat}{3}
& \varphi_{n}(x,t)&&=A^{-1}(x,t)\Phi\left(\frac{\mu x+\gamma}{\sigma}\right) \\
& &&=\exp\left\{-\left(\frac{\mu^{2}}{\sigma^{2}}+i\frac{\mu\mathcal{W}_{\mu}}{\sigma}\right)\frac{x^2}{2} - \left(\frac{\mu\gamma}{\sigma^{2}}+i\frac{\mu\gamma}{\sigma^2}\right)x-\left(\frac{\gamma^{2}}{2\sigma^{2}}+i\frac{\mu\xi}{\sigma}\right) \right\} \\
& && \hspace{40mm}\times \sqrt{\frac{1}{2^{n}n!\sqrt{\pi}}} \sqrt{\frac{\mu}{\sigma}} H_{n}\left(\frac{x+\gamma}{\sigma} \right) \, .
\end{alignedat}
\label{eq:INN3}
\end{equation}
Notice that the orthogonality condition obtained from~(\ref{eq:INN2}) holds provided that both solutions are evaluated at the same time, that is, $\langle\psi_{n}(t)\vert\psi_{m}(t)\rangle=\delta_{n,m}$. In turn, the orthogonality can not be taken for granted at different times, $\langle\psi_{m}(t')\vert\psi_{n}(t)\rangle\neq\delta_{n,m}$ for $t\neq t'$. Moreover, the space of solutions generated with~(\ref{eq:INN2-1}) is dynamic, $\mathcal{H}(t)=\operatorname{Span}\{\vert\psi_{n}(t)\rangle \}_{n=0}^{\infty}$. Such a property is beyond the scope of this work and will be studied elsewhere. For information on the matter see~\cite{Ali18}.

We have shown the orthonormality of the solutions $\psi_n(x,t)$, however, it is necessary to emphasize that they are not eigenfunctions of the Hamiltonian $\hat H(t)$. Nevertheless, the functions $\psi_n(x,t)$ are admissible from the physical point of view. Since $\hat H(t)$ is not a constant of motion of the system $\frac{d}{dt}\hat{H}(t)\not=0$, we wonder about the observable that define the system uniquely. Such observable must include the set $\{ \psi_n(x,t)\}_{n=0}^{\infty}$ as its eigenfunctions. Moreover, what about the related spectrum? The latter points must be clarified in order to provide the functions (\ref{eq:INN2-1}), and any linear combination of them, with a physical meaning.

Remarkably, such information is obtained from the point transformation itself, because any conserved quantity is preserved \cite{Ste93}. Indeed, from \eqref{eq:INT3} we see that the energy eigenvalues $E_{n}=(n+1/2)$ of the stationary oscillator must be preserved since they are constant quantities. To be specific, using the relationships~\eqref{eq:INT10} together with $A(x,t)$=(\ref{eq:INT15}), the stationary eigenvalue equation~\eqref{eq:INT2-2} is then deformed into the new eigenvalue equation
\begin{multline}
-\frac{\sigma^2}{2\mu^{2}}\frac{\partial^2\varphi_{n}}{\partial x^2}+\frac{1}{2}\left( \mathcal{W}_{\mu}^{2}+\frac{\mu^{2}}{\sigma^2} \right)x^2 \varphi_{n} - \frac{\sigma\mathcal{W}_{\mu}}{2\mu}i\left(2x\frac{\partial}{\partial x} + 1 \right)\varphi_{n} \\
- \frac{\sigma \mathcal{W}_{\gamma}}{\mu}i\frac{\partial\varphi_{n}}{\partial x} +\left(\mathcal{W}_{\mu}\mathcal{W}_{\gamma}+\frac{\mu\gamma}{\sigma^{2}} \right)x \varphi_{n} +\frac{1}{2}\left(\mathcal{W}_{\gamma}^{2}+\frac{\gamma^2}{\sigma^2}\right) \varphi_{n} = E_n \varphi_{n},
\label{eq:INV1}
\end{multline}
where the eigenvalues $E_{n}=(n+1/2)$ have been inherited from the stationary oscillator. It is immediate to identify the operator
\begin{multline}
\hat{I} (t)=\frac{\sigma^2}{2\mu^{2}}\hat{p}_{x}^2+\frac{1}{2}\left(\mathcal{W}_{\mu}^{2}+\frac{\mu^{2}}{\sigma^2}\right)\hat{x}^2+ \frac{\sigma\mathcal{W}_{\mu}}{2\mu}(\hat{x}\hat{p}_{x}+\hat{p}_{x}\hat{x})+\frac{\sigma W}{\mu} \hat{p}_{x} \\ 
+\left( \mathcal{W}_{\mu}\mathcal{W}_{\gamma}+\frac{\mu\gamma}{\sigma^{2}} \right)\hat{x} + \frac{1}{2}\left(\mathcal{W}_{\gamma}^{2}+\frac{\gamma^2}{\sigma^2} \right) \mathbb{I}(t), 
\label{eq:INV2}
\end{multline}
with $\mathbb I(t)=\sum_{n=0}^{\infty}\vert\psi_{n}(t)\rangle\langle\psi_{n}(t)\vert$ the representation of the identity operator in ${\cal H}(t)$. The invariant operator $\hat{I}(t)$ is such that we get the eigenvalue equation 
\begin{equation}
\hat{I} (t)\vert\varphi_{n}(t) \rangle = (n+1/2)\vert\varphi_{n}(t)\rangle .
\label{eq:INV1-1}
\end{equation}
Besides, straightforward calculations show that $\hat{I}(t)$ satisfies the invariant condition 
\begin{equation}
\frac{d}{dt}\hat{I}  (t)=i[\hat{H}(t),\hat{I} (t)]+\frac{\partial}{\partial t}\hat{I}(t)=0.
\label{eq:INV3}
\end{equation}
That is, $\hat{I}(t)$ is an integral of motion of the parametric oscillator. Remark that the invariant operator $\hat{I}(t)$ arises in natural way from the point transformation, without necessity of any ansatz as in~\cite{Lew69}. On the other hand, with $\hat{I}(t)$ and~\eqref{eq:INN2-1} we find
\begin{equation}
\psi_{n}(x,t)=e^{-i\hat{I} (t) \tau(t)} \varphi_{n}(x,t)=e^{-iw(n+1/2) \tau (t)}\varphi_{n}(x,t).
\label{eq:INV3-1}
\end{equation}
Thus, we can conclude that the time-dependent complex-phase of the Lewis and Riesenfeld approach~\cite{Lew69} coincides with the exponential term in~(\ref{eq:INV3-1}), that is, such phase is proportional to the deformed time parameter $\tau(t)$. Notice that, contrary to the stationary case, the operator $e^{-i\hat{I} (t) \tau (t)}$ in~\eqref{eq:INV3-1} is not the time evolution operator. 

In particular, for $\gamma_{1}=\gamma_{2}=F(t)=0$ and a constant mass $m(t)=\mu(t)=1$, the operator (\ref{eq:INV2}) coincides with the invariant of Lewis and Riesenfeld~\cite{Lew69}.

%%%%%%%%%%%%%%%%
%\begin{figure}
%\centering
%\begin{tikzpicture}
%  \matrix (m) [matrix of math nodes,row sep=5em,column sep=8em,minimum width=2em]
%  {
%     i\frac{\partial}{\partial \tau}\Psi=\hat{H}_{osc}\Psi
%     & i\frac{\partial}{\partial t}\psi=\hat{H}(t)\psi \\
%     \hat{H}_{osc}\Phi_{n}= (n+1/2)\Phi_{n} 
%     & \hat{I}(t)\varphi_{n}=(n+1/2)\varphi_{n} \\};
%  \path[-stealth]
%    (m-1-1) edge [<->,very thick,red] node [left] {$\Psi_{n}=e^{-i\hat{H}_{osc}\tau}\Phi_{n}$} (m-2-1)
%    (m-2-2) edge [<->,very thick,red] node [right] {$\psi_{n}=e^{-i\hat{I}_{2}\tau(t)}\varphi_{n}$} (m-1-2)
%   	(m-1-1) edge [blue] node [above] {\textcolor{black}{P.T.}} (m-1-2)
%   	(m-1-1) edge [->,very thick,blue] node [below] {\textcolor{black}{$y(x,t)$, $\tau(t)$, $\psi=A(x,t)\Psi$}} (m-1-2)
%   	(m-2-1) edge [->,very thick,blue] node [below] {\textcolor{black}{P.T.}} (m-2-2);
%\end{tikzpicture}
%\caption{\footnotesize Connection between the stationary and parametric oscillators through the point transformation (P.T. for short). The orientation of the blue (horizontal) arrows may be inverted with the construction of the inverse point transformation. Thus, the diagram is commutative.}
%\label{fig:DIA}
%\end{figure}
%%%%%%%%%%%%%%%%

%---------------------------------------> Subsection

\section{Concluding remarks}
It was shown that a set of orthonormal solutions for the time-dependent mass nonstationary oscillator can be found by constructing the appropriate point transformation and deforming the solutions of the stationary oscillator. The latter is possible since the point transformation preserve the structure of the inner product. Although the Hamiltonian depends explicitly on time, an spectral problem can be found for the appropriate constant of motion which emerges as the transformed Hamiltonian of the stationary oscillator. The procedure has been developed in general for any time-dependent mass, frequency and an external driving force. Among the examples that could be addressed we have the Caldirola-Kanai oscillator, which leads to the quantum Arnold transformation~\cite{Gue12}, and the Hermite oscillator~\cite{Una18}. These and some other models will be discussed in detail elsewhere.
\subsection*{Acknowledgment}
K. Zelaya acknowledges the support from the Mathematical Physics Laboratory, Centre de Recherches Math\'ematiques, through a postdoctoral fellowship. He also acknowledges the support by Consejo Nacional de Ciencia y Tecnolog\'ia (Mexico), grant number A1-S-24569. V. Hussin acknowledges the support of research grants from NSERC of Canada.

\end{document}